# BEYOND LIKE-FOR-LIKE: A USER-CENTERED APPROACH TO MODERNIZING LEGACY APPLICATIONS

M. Polzin*, M. Guzman†, Fermi National Accelerator Laboratory, Batavia, USA

*Abstract*

When modernizing a legacy application, it is easy to fall back on a like-for-like replica with new tools and updated design stylings, but this is an opportunity to explore making a more intuitive application that supports user tasks and efficiency. Rather than having a blank canvas–unburdened by legacy tech debt–to create a new application, you are working with an existing application that is integral to accelerator operations and one that expert users are already familiar with. Due to this, you might assume people will prefer the like-for-like, but you could be carrying forward the pain points, processes that are inefficient, and ultimately wind up with an application that no one wants to use because it doesn't solve existing problems.

Getting users involved can make all the difference in your approach to modernizing a legacy application that caters to both newer and expert users. It also can bridge the gap between like-for-like and introducing new GUI design. Having a legacy application doesn't have to make the modernized one difficult to develop, as the existing application is a tool in how you move forward with the new application. It provides insight into areas that an application with a clean slate doesn't give you.

## INTRODUCTION

At Fermilab, the Accelerator Controls Operations Research Network (ACORN) project is working to modernize Fermilab's legacy particle accelerator control system [?]. Within this larger project, our team is tasked with modernizing the existing control system applications and increase usability by creating intuitive and effective user interfaces. We are emphasizing user experience efforts by incorporating user input throughout our process. Although we could take a like-for-like approach and rebuild existing applications with newer technology, we have decided to seek opportunities for improvements to better meet the needs of our users.

## MODERNIZING LEGACY APPLICATIONS

Fermilab has over 600 legacy applications that are a combination of text based C/C++ applications and Java applications. A goal of the ACORN project is to modernize existing applications by providing a web browser-based interface to new applications that provide the functionality of existing legacy applications. In order for ACORN to modernize all existing applications, a full assessment of legacy applications must be done. The initial assessments were to inventory the lines of code for each application and look at usage statistics over a span of two years. Having the lines of code allowed to have a high level understanding of the complexity of each application and having usage statistics allowed to prioritize higher used applications.

### Like-for-Like Pitfalls

When modernizing applications one strategy could be to develop a like-for-like replacement. This strategy would take an existing application and replicate all functionality as well as the interface. This could be short term success as the development cycle could be quick. However, this strategy could be filled with pitfalls. A like-for-like replacement would carry over the user interface of legacy applications–which includes structural design choices and user workflows. This could mean that existing pain points from users will continue to exists. This strategy also does not take into account that the design decisions were based on the limits and strengths of technologies that are no longer being used. While the application will be modernized, the experience would not be.

### Moving Beyond Like-for-Like

Another strategy would be moving away from a like-for-like application. This allows for a reevaluation of the needs of users and purpose of the application, as those may have changed since the development of the initial application. A reevaluation of the application allows the development to leverage the strengths of current technologies and address inefficiencies in the user workflows. For example, moving to a browser-based application allows for leveraging the built in usability features of browsers, cross platform usage, and responsive experience. This can introduce an increase in usability and a reduction in learning curves. By moving to modern technologies you can also move away from technical debt and allow for adaptability and future improvements.

## LEGACY APPLICATIONS AND USERS

Modernizing legacy applications may feel constraining in comparison to applications that do not yet exist where you have more freedom, but there are ways to utilize the legacy applications and find opportunities for improvements in efficiency and usability.

Learning from users how they use the legacy applications helps define tasks and goals as well as what they do with the information they engage with. You can focus on user workflows within the application to ensure you are meeting their needs that may not be addressed in a one-for-one replica. You may also find certain tasks are completed by interacting with multiple applications. By investigating the relationships between applications and tasking, you can uncover redundancies and organize information in a way that better fits the user needs.

---

* mpolzin@fnal.gov
† mguzman@fnal.gov

When interacting with users and the legacy applications, it is important to work with both novice and expert users. Expert users know the control system very well (and the shortcuts). For example, if you only talk to expert users, you increase the chances of replicating an application that may not only be less intuitive for newer users, but also carries over any time consuming learning curve that may exist. At Fermilab, we found that new operators are not confident using certain control system applications. A like-for-like replacement would not meet the needs of novice users.

## USER RESEARCH

When moving from an existing application, user experience efforts may be viewed as something that is incorporated only at the design and development stages. Getting user input while building is extremely important, but also important is understanding the users that the application is built for. Although you may already be familiar with your users due to the original application, it can be beneficial to spend time with your users before proceeding with building the new application. In order to ensure that the application will address user needs and support user tasks, it is important to understand why and how they use the application as well as how it fits into their workflow to achieve goals and tasks.

### Know Your Users

A foundation of user knowledge can help align the application with user needs. It may be easy to jump into an application and build from the functionality and layout of the existing application, but with this lies potential segmented results due to passing the opportunity to see the bigger picture and the story the application tells. To simplify, if you do not understand who the users are and why they use or need an application, you potentially run into the problem of designing for the application and not the user.

Rather than "this application needs to have this," it becomes "the user needs to be able to do this." By approaching an application from this angle, there is the space for ideating efficient and intuitive solutions that better assist the user.

### Novice vs. Expert Users

Novice and expert users often have different needs. Novice users may need additional context, whereas expert users may have ample experience with the legacy applications and need shortcuts. If you gather information specifically from expert users, you could find yourself with an application that is not intuitive to novice users and lacks context needed for a novice user. If you gather information specifically from novice users, you could wind up with an application that expert users find frustrating. Thus, it is important to get input from both user types.

### User Research Methods

There are many different user research methods that can be used for to learn about your users. A powerful method for learning about your user is performing user interviews in which you can ask questions directly. An important thing to note with user interviews (and gathering user input in general), is that you do not want to "lead" with your questions. Asking leading questions can compromise the validity of the information gathered [2].

Observation is another powerful tool in which you do not engage, but see how users use an application. The way in which someone talks about a task may differ from what is done in practice.

## USER INVOLVEMENT

Once you have a strong foundation of user knowledge, the way in which you interact with users can evolve to align with application development. As you design and develop the application, users can be included to collect requirements, review user interface designs, and provide input on the application as it comes to fruition.

At the design and development stages, user input can help refine an application and ensure that the new application will not only meet user needs, but also improve their experience completing their associated tasking with the application. Now user interviews will be less essential whereas usability testing and user input sessions will be valuable. It is important to continually have users engaged to avoid spending time on an application that once deployed, does not meet the user needs and/or is rejected by the users.

Wireframes (blueprints of the basic application structure), mock-ups, and interactive prototypes can be useful tools when collecting user input. In a modernization from an existing application, you can collect information on the existing application and then test those scenarios/use cases in the new application and measure efficiency. Figure 1 shows the simplicity of a wireframe.

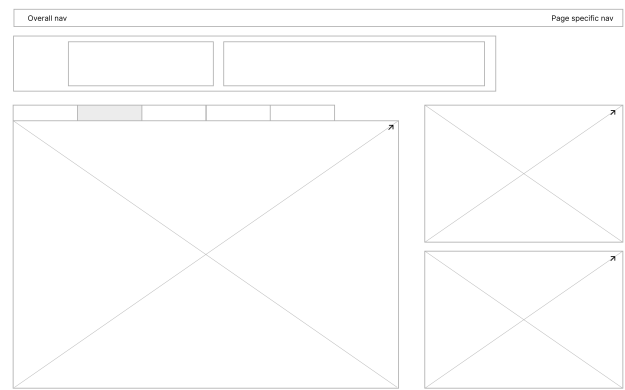

Figure 1: Example wireframe of a control system application

### User Feedback

The term "feedback" can be often be seen as asking someone's preference. Perhaps you collect feedback and the user responds with comments about the look and feel of the application. It is important to gather more information from the user, grounded in the user's task demands and their goals.

## Scenario Testing

When performing usability testing with users, you have the opportunity to verify use cases and workflows as well as weed out pain points and any elements that are confusing or need improvement. One way to assist in gathering this input, is through scenario testing. Scenario testing can be done with both functional prototypes as well as active developed builds of the application. In order to do this, the user is given a task to complete with the application. The scenario can be a use case that was captured when you collected input on user needs. For example, retrieve an old data set and plot a waveform.

If the task is present in both the legacy application and modernized application, users can partake in scenario testing with both applications. This will help gather insights on efficiency and success of the user workflow solution.

## CASE STUDIES

As we move through the modernization of our control system applications, we have had the opportunity to interact with users and include them in our approach as we work to reassess and improve legacy applications. At times we are verifying existing workflows, addressing issues and pain points, and collaborating to ensure an application meets both the needs of a novice user and expert user.

### Autotune

We worked on modernizing the Autotune application. The purpose of the Autotune application is to monitor beam trajectory in transfer lines and make corrector adjustments when the trajectory has moved away from the desired trajectory. The current Autotune application (shown in Fig. 2) is a Java application that uses a tomcat server to do the calculations and initiates the corrector changes. The Main Control Room will have an instance of the Autotune client application for each beamline running at all time. The Autotune server is also responsible for generating alarms for various statuses such as when Autotune is disabled or the BPM readings are out of position.

When we initiated gathering requirements, we focused on Main Control Room Operators as they are responsible for monitoring that the Autotune application is applying the appropriate corrector changes. Through user engagement we found that while the application does not need to be open for operation, the control room Operators preferred to keep the application open at all times. The Operators informed us that having the trajectory plots open was a way for Operators to verify that the Autotune was functioning correctly. They mentioned that often the alarms associated with the Autotune application would not get generated at the appropriate time and they would not rely on the alarms. This led us to believe that the plot on the main page of the application was not necessary if the alarms could be configured to accurately notify Operators when they need to engage.

However, speaking with other users informed us that their workflows were made easier by using the plot, thus having it on the main page would be helpful. Beamline physicists and machine specialists mentioned that being able to see the trajectory through the transfer line allowed them to understand the beam quality as it was injected into the synchrotron. This helped them understand any losses generated and they were able to understand the causes. Whilst engaging with the physicists and specialists, we found pain points and opportunities to improve the plots to better help their tasking.

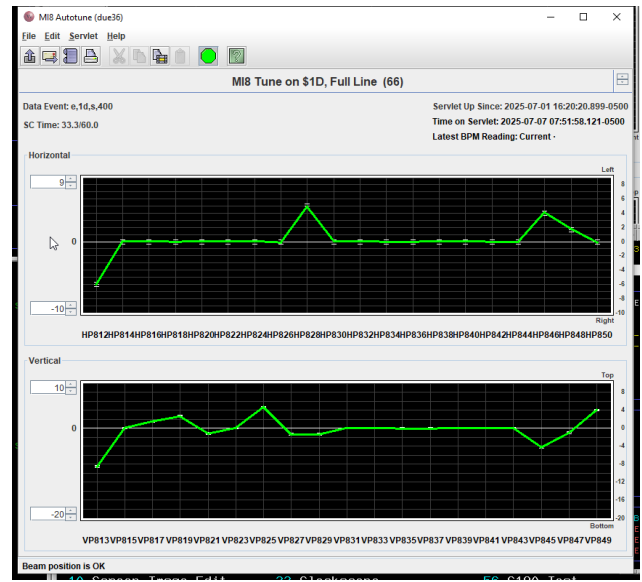

Figure 2: A screenshot of the Autotune application running in the main control room.

Through user engagement, we were able to understand the needs of users. While we focused on Main Control Room Operators, it was important to engage with a variety of users to understand all use cases. This engagement with of variety of users allowed us to discover pain points and help build an application that addresses the needs of all users, not just the primary audience.

### Beam Position Monitor (BPM)

The BPM application (Fig. 3) triggers a plotting window of beam position monitors throughout an accelerator or beamline. It allows the user to define points in the cycle they are interested in. This particular application is outside of ACORN's scope; however, we were brought in to help design and align with ACORN effort. This project is ongoing and not yet finalized.

To begin, we met with users to gather requirements from the existing application. We observed how users move through the application and found that users jumped around to different sections to set up a plot. We also noticed there is a lack of clarity in regard to the lack of contextual clues. This places a cognitive load on users to remember what is required to interact with as they parse through the sections to find the relevant content. Thus when we went to design the new application, we made sure to capture this information to ensure there was an obvious flow as a user moves through the application. Through iteration, we were able to streamline the plot configuration process in which the user can follow a clear path that supports the common use cases that would bring a user to the application. What once required the user to interact with different areas of the page, they now have a clear path that helps with ease.

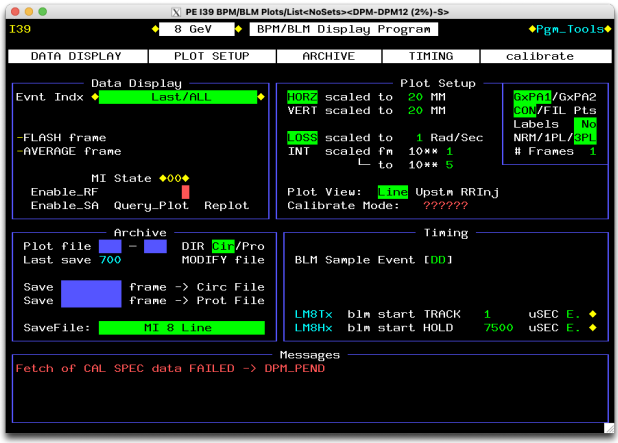

Figure 3: The legacy BPM application.

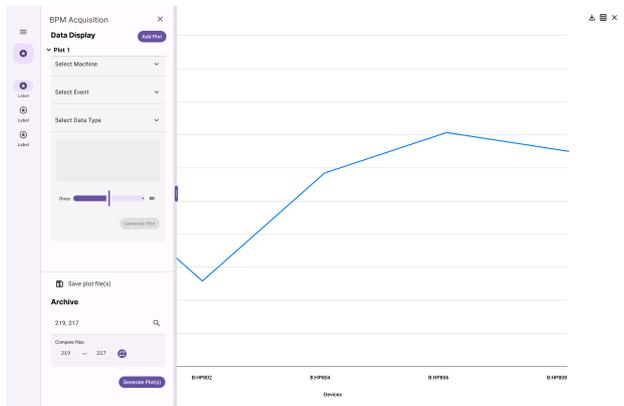

Figure 4: An early prototype for the BPM application.

Through different stages of wireframing and prototyping, user input was collected to help refine the user interface design. Seeing how a user moves through a functional prototype helped fix potential pitfalls that reduced clarity or were less intuitive upon interaction. The application has grown into a more efficient, clear, and intuitive interface that addresses current pain points and features new solutions that enrich the user's experience, as shown in Fig. 4.

Had the application been a like-for-like replacement, we would have carried over a design that covers the necessary functionality, but requires expert knowledge and places a heavy cognitive workload on the user.

## CONCLUSION

Like-for-like replacements may seem like the quick and easy solution, but that approach does have its pitfalls. If existing problems are not addressed, then the same issues and inefficiencies are carried over to the newer application. A modernization effort gives the opportunity to increase usability, meet any new needs that developed since the legacy application was developed, and improve the user workflows.

Incorporating user input early on can save time and effort, and can help with user acceptance by developing an application that the user wants to use. You can utilize what you already have with the legacy applications to learn areas of improvement and bridge the gap between the old and new applications whilst evolving.